\documentclass[useAMS,usenatbib]{mn2e}

\usepackage{graphicx}
\usepackage{lscape}
\usepackage{deluxetable}
\bibpunct{(}{)}{;}{a}{}{,}

\title[The link between the optical continuum and the jet of the
3C\,390.3]{Observational evidence for the link between the variable
optical continuum and the subparsec-scale jet of the radio galaxy 3C\,390.3}

\author[T.G.Arshakian et al.]  {T.G. Arshakian$^{1}$\thanks{E-mail:
tigar@mpifr-bonn.mpg.de (TGA)},
J. Le\'on-Tavares$^{1,2}$, A.P. Lobanov$^{1}$, V.H. Chavushyan$^{2}$,
\\ \\ {\rm\LARGE A.I. Shapovalova$^{3}$, A.N. Burenkov$^{3}$ and J.A. Zensus$^{1}$} \\ \\
$^{1}$Max-Planck-Institut f\"ur Radioastronomie, Auf dem H\"ugel 69,
53121 Bonn, Germany\\ $^{2}$Instituto Nacional de Astrof\'{\i}sica
\'Optica y Electr\'onica, Apartado Postal 51 y 216, 72000 Puebla, Pue,
M\'exico\\ $^{3}$Special Astrophysical
Observatory of the Russian AS, Nizhnij Arkhyz, Karachaevo-Cherkesia
369167, Russia}

\begin{document}

\date{Accepted ??? December 15. Received ??? December 14; in original form ??? October 11}

\pagerange{\pageref{firstpage}--\pageref{lastpage}} \pubyear{2002}

\maketitle

\label{firstpage}

\begin{abstract}
The mechanism and the region of generation of variable continuum emission are poorly understood for radio-loud AGN because of a complexity of the nuclear region. High-resolution radio VLBI (very long baseline interferometry) observations allow zooming into a subparsec-scale region of the jet in the radio-loud galaxy 3C\,390.3. We combined the radio VLBI and the optical data   covering the time period of 14 years to look for a link between optical flares and parsec-scale jet. We identify two stationary and nine moving radio features in the innermost subparsec-scale region of the jet. All nine ejections are associated with optical flares. We found a significant correlation (at a confidence level of $>99.99\%$) between the ejected jet components and optical continuum flares. The epochs at which the moving knots pass through the location of a stationary radio feature and the optical light curve reaches the maximum are correlated.
The radio events follow the maxima of optical flares with the mean time delay of $0.10\pm0.04$ years. This correlation can be understood if the variable optical continuum emission is generated in the innermost part of the jet. A possible mechanism of the energy release is the ejection of knots of high-energy electrons that are accelerated in the jet flow and generate flares of synchrotron continuum emission in the wide range of frequencies from radio to X-ray bands. In this scenario,
the beamed optical continuum emission from the jet and counterjet ionizes a gas in a subrelativistic outflow surrounding the jet, which results in a formation of two outflowing conical regions with broad emission lines (in addition to the conventional broad line region around the central nucleus).
\end{abstract}


\section[]{Introduction}
The variable continuum flux in AGN, signaling the activity of the
central engine, is detected throughout the entire electromagnetic
spectrum, on time-scales from days to years
\citep{peterson02,zheng,wamsteker,shapo}. The bulk of the continuum flux is
believed to be generated in the accretion disk and is
responsible for ionizing the cloud material in the
broad-line region (BLR). Localization of the source of the variable
continuum emission in AGN is instrumental for understanding the
mechanism for release and transport of energy in active galaxies. In
radio-quiet AGN, representing about 90\,\% of the AGN population,
the presence of rapid X-ray flux variations and iron emission line
(Fe K$\alpha$) indicates that most of the soft X-ray emission
originates from the accretion disk \citep{mushotzky93}.  In
radio-loud AGN, the activity of the central engine is accompanied by
highly-relativistic collimated outflows (jets) of plasma material
formed and accelerated in the vicinity of the black hole
\citep{ferrari98}. Inhomogeneities in the jet plasma appear as a
series of compact radio knots (jet components) observed on scales
ranging from several light weeks to about a kiloparsec
\citep{alef,kellermann04}.

\begin{figure}
\resizebox{\hsize}{!}{\includegraphics[angle=-90]{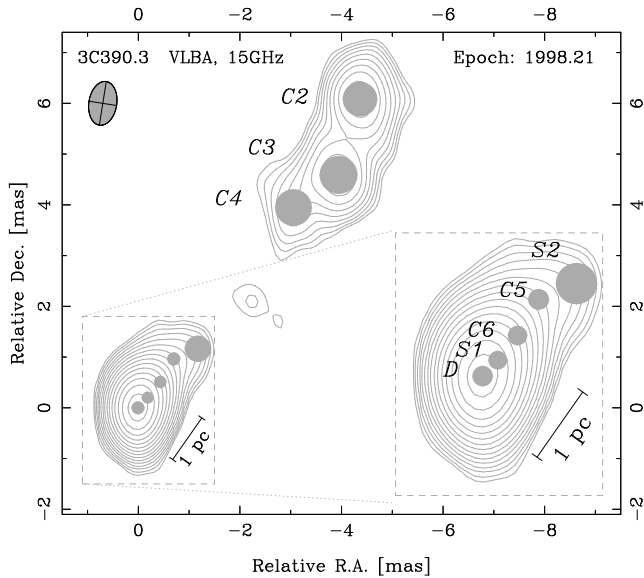}}
\caption{Radio structure of 3C 390.3 observed in 1998.21 with very
long baseline interferometry at 15\,GHz (2\,cm). Innermost fraction of
the jet is shown in the inset.  The resolving point-spread function
(beam) plotted in the upper left corner is $0.87 \,\mathrm{mas} \times 0.55$ mas
oriented at an angle of $8^\circ$ (clockwise rotation).  The peak
flux density in the image is 190\,mJy/beam ($3.1\times 10^9$\,K) and
the rms noise is 0.2\,mJy/beam.
The contours are drawn at $1,\,\sqrt{2},\,2\,...$ of the lowest
contour shown at 0.6\,mJy/beam.
The labels mark three stationary features (D, S1 and S2) and a
subset of moving components (C2--C6) identified in the jet. Note that
six more components, C7-C12, have been first identified in the jet
in the VLBA images at later epochs (see Fig.~\ref{rfit}).}
\label{rmap}
\end{figure}

The unification scheme \citep{urry95} of radio-loud AGN suggests that the optical continuum associated with the accretion flows and broad emission lines are viewed directly in radio-loud quasars and BL Lacertae objects (BL Lacs). The Doppler boosted continuum emission from the
relativistic jet may dominate at all energies in BL Lacs \citep[see][]{ulrich,worrall},
suggesting that the continuum variability in radio-loud AGN is related to
both the jet and the instabilities of accretion-disk flows
\citep{mushotzky93,ulrich}. In radio-loud quasars, the beaming of continuum emission is less but optical synchrotron radiation from the core of the jet \citep[3C\,273;][]{soldi08} can be still significant. In radio-loud galaxies the emission from accretion disk or hot corona can be hidden by an obscuring torus, and hence, the bulk of the optical continuum and broad-line emission in radio galaxies can be attributed to the relativistic jet rather than the central
engine if the boosted jet-emission is still strong, i.e. the jet is intrinsically luminous and highly relativistic.

The presence of a positive correlation between beamed radio
luminosity of the jet and optical nuclear luminosity in the sample
of radio galaxies suggests that the optical emission is non-thermal
and may originate from a relativistic jet
\citep{chiaberge99,chiaberge02,hardcastle00}. In radio-loud quasars, there is evidence for a variable synchrotron optical flare generated around the radio core of PKS\,0420-014 \citep{darcangelo07} and in the innermost $\approx0.2$\,pc region of the jet in BL Lac \citep{marscher08}. A correlation between ejection of superluminal knots and dips in the X-ray emission was reported for the radio galaxy 3C\,120 \citep{marscher02}. Evidence for correlated radio and optical variations has been reported for several AGN with time lags up to few hundreds days \citep{hanski02}. The detection of a
correlation between variability of the
continuum flux and changes in the radio structure in a radio galaxy on subparsec-scales would be the most direct
evidence of optical continuum emission coming from the jet. We combine the
results from monitoring of the double-peaked broad line radio-loud galaxy
3C\,390.3 (redshift $z=0.0561$) in the optical from 1992 to 2008 time period
\citep[][ Shapovalova et al, in preparation]{shapo,sergeev}, UV \citep{zheng}, and X-ray
\citep{leighly,gliozzi06} regimes with 21 VLBI observations of its radio emission made from
1994 to 2008 as part of the 15\,GHz VLBA\footnote{Very Long Baseline
Array of National Radio Astronomy Observatory, Socorro, NM, USA}
survey \citep{kellermann04} and our VLBA observations during the time
period from 2005 to 2008.

\begin{figure}
\resizebox{\hsize}{!}{\includegraphics[angle=90]{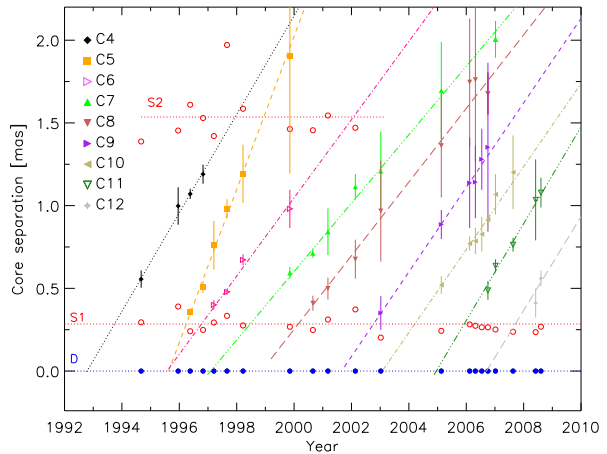}}
  \caption {Separation of the jet components relative to the
  stationary feature D (blue circles) for 21 epochs of VLBI
  observations. The stationary components S1 and S2 are marked by open red circles.
  Designations of moving components are shown in the upper left corner.
  The lines represent the
  best linear least-squares fits to the component separations.}
  \label{rfit}
\end{figure}

In Sect.~\ref{sec:jet_structure}, we analyze the structure and
kinematics of the pc-scale jet in 3C\,390.3. Correlations between
properties of the compact jet and nuclear optical emission on
scales less than one parsec are analyzed in Sect.~\ref{sec:link}.
Variable radio emission from stationary components of the jet is discussed in Sect.~\ref{sec:var_emission}, and the identification of these stationary components is discussed in Sect.~\ref{sec:id}. In Sect.~\ref{sec:discussion}, we discuss possible radiation mechanism acting in the nuclear region of 3C\,390.3.

\begin{table*}
  \label{table:jet_kinematics}
  \begin{center}
  \begin{minipage}{140mm}
   \caption{Kinematics and ejection epochs of moving jet components in 3C\,390.3.}
   \begin{tabular}{lccccc}
   \hline

      Comp. & $t_{\rm D}$ & $t_{\rm S1}$ & $\mu_\mathrm{r}$ & $\beta_\mathrm{app}$ \\
            &  (yr)       &    (yr)      & (mas\,yr$^{-1}$) &                      \\
       (1)  &  (2)        &    (3)       & (4)              &  (5)                 \\
      \hline

   C4  &  1992.79$_{-0.46}^{+0.37}$  &  1993.70$_{-0.35}^{+0.28}$   & 0.29$\pm$0.03 &  1.05$\pm$0.11\\
   C5  &  1995.62$_{-0.08}^{+0.07}$  &  1996.21$_{-0.04}^{+0.03}$   & 0.46$\pm$0.03 &  1.63$\pm$0.11\\
   C6  &  1995.66$_{-0.22}^{+0.18}$  &  1996.75$_{-0.09}^{+0.07}$   & 0.24$\pm$0.02 &  0.87$\pm$0.09\\
   C7  &  1997.15$_{-0.47}^{+0.36}$  &  1998.43$_{-0.28}^{+0.22}$   & 0.21$\pm$0.03 &  0.74$\pm$0.09\\
   C8  &  1999.29$_{-0.36}^{+0.26}$  &  2000.38$_{-0.15}^{+0.10}$   & 0.22$\pm$0.04 &  0.78$\pm$0.13\\
   C9  &  2001.65$_{-0.74}^{+0.52}$  &  2002.71$_{-0.51}^{+0.39}$   & 0.26$\pm$0.05 &  0.90$\pm$0.16\\
  C10  &  2003.11$_{-0.49}^{+0.36}$  &  2004.17$_{-0.29}^{+0.21}$   & 0.25$\pm$0.04 &  0.90$\pm$0.14\\
  C11  &  2004.89$_{-0.56}^{+0.37}$  &  2005.84$_{-0.32}^{+0.21}$   & 0.28$\pm$0.06 &  1.01$\pm$0.21\\
  C12  &  2006.63$_{-1.49}^{+0.79}$  &  2007.73$_{-0.64}^{+0.34}$   & 0.28$\pm$0.08 &  1.00$\pm$0.27\\
\hline

\end{tabular}


\medskip
\begin{flushleft}
Columns are as follows:
(1) -  Component identifier,
(2) -  Ejection time from  the component D,
(3) -  Passing time through the stationary component S1,
(4) -  Angular radial speed,
(5) -  Radial speed in units of the speed of the light.
\end{flushleft}

\end{minipage}
\end{center}
\end{table*}

\section{Structure and kinematics of the parsec-scale jet}
\label{sec:jet_structure}
To parameterize the structure of the radio emission, we applied the
technique of modelfitting \citep{pearson} and fit interferometric
visibility amplitudes and phases from each of the 21 VLBA datasets by a set of two-dimensional, circular Gaussian features (shaded
circles in Fig.~\ref{rmap}). Similar fits have been obtained for 17 observations
from the 15\,GHz VLBA survey database \citep{kellermann04} and for
four our observations with the VLBA at 15\,GHz, for the
purpose of cross-identifying and tracing different features in the
jet (Table~\ref{tbl-1} in the Appendix~A).
We use the positions and flux densities of these
components for tracing the evolution of the jet emission on angular
scales of $\sim 3$ milliarcseconds (mas). At the distance of
3C\,390.3, 1\,mas corresponds to a linear distance of 1.09\,pc
for the flat $\Lambda$CDM cosmology with the Hubble constant $H_0 =
70$\,km\,\,s$^{-1}$\,Mpc$^{-1}$ and the matter density
$\Omega_\mathrm{m} = 0.3$.

\begin{figure*}
\resizebox{\hsize}{!}{\includegraphics[angle=-90]{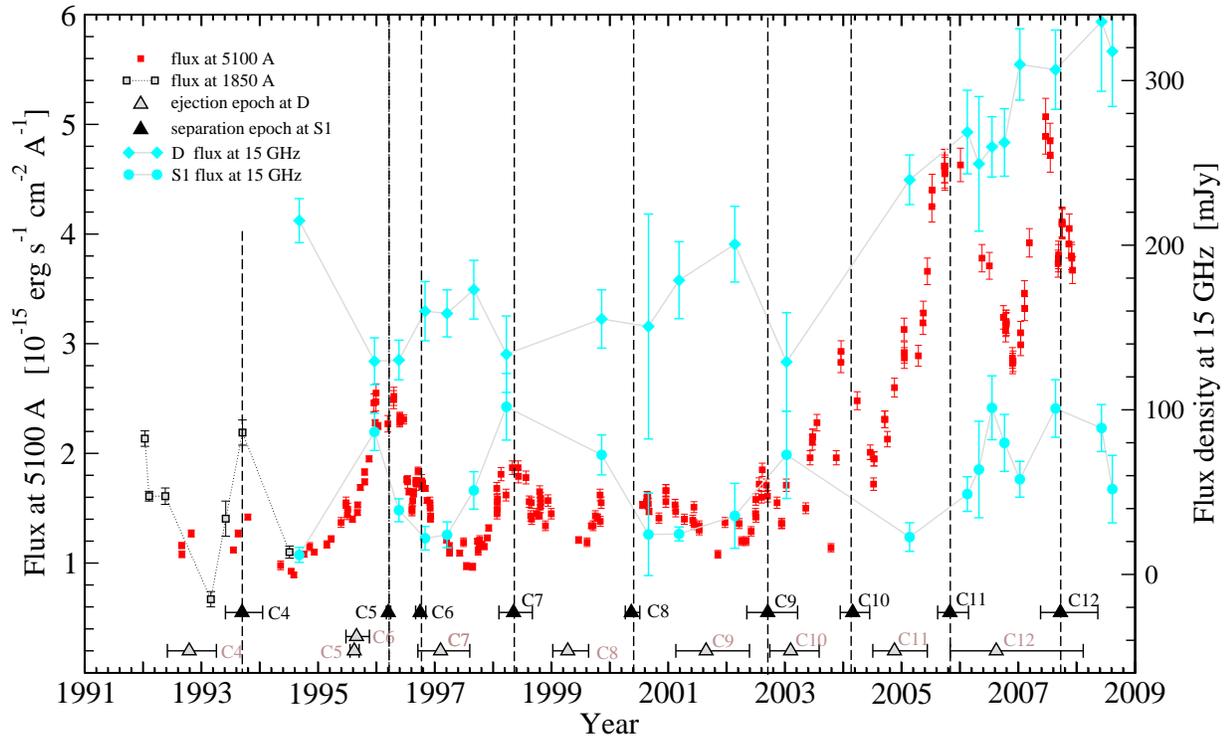}}
  \caption {The variations of the continuum fluxes
  of 3C 390.3 during the 1992 to 2008 time period. Optical continuum
  light curve at 5100\AA\
  \citep[red squares; ][]{shapo,sergeev}, UV
  continuum fluxes \citep{zheng} scaled by a factor of 50 (grey
  squares), and variations of radio flux density from D and S1
  components (filled cyan diamonds and circles) are presented. The
  times of ejection, $t_\mathrm{D}$, of radio knots from D and the 
  times of their separation, $t_\mathrm{S1}$,
  from the component S1 are marked by open triangles and filled
  triangles respectively.}
\label{flux-epoch}
\end{figure*}

Based on the separations, sizes and flux densities of the Gaussian components
located within 3\,mas of the feature
D at the narrow end of the jet (Fig.~\ref{rmap}), we identified nine
moving components C4--C12 (in addition to the previously known
components C2 and C3; Alef et al. 1996) and two stationary features S1
and S2 separated from D by $(0.28\pm0.03)$\,mas and $(1.50\pm0.12)$\,mas,
respectively (Fig.~\ref{rfit}). Note that two innermost components D and S1 are
spatially resolved and their flux densities are uncorrelated (see Appendix~A).
The more distant stationary feature S2
is probably related to a small change in the direction of the flow that
causes the relativistic brightening of the radio emission
\citep{gomez97}. The presence of S1 may be related to both geometrical
and physical factors. Recent results from close VLBI monitoring of a
sample of parsec-scale jets \citep{jorstad01} imply that such
stationary features are likely to be common in compact relativistic
flows.

Proper motions of (0.2 to 0.5)\,mas/yr (which
correspond to apparent speeds of (0.7 to 1.6)\,$c$, where $c$ is the
speed of light) are estimated from
linear fits to the observed separations of the components C4--C12
from the component D (Table~\ref{table:jet_kinematics}). Resolution of the VLBA data at 15\,GHz is not
sufficient for detecting moving components at locations between
D and S1, since D and S1 are strong, unresolved features separated
by less than one beamwidth, while the moving features are less
prominent and more extended. High frequency VLBI observations are
required to study these angular scales in 3C\,390.3.
Back-extrapolation of the fits for C5 and C6
indicates that these two components may originate from a single
event. They could result from a moving perturbation in the jet that
creates a forward and a reverse shock pair \citep{gomez97} or
trailing features behind a strong relativistic shock
\citep{gomez01,agudo01}. We use the linear fits to estimate, for
each moving component, the epoch, $t_\mathrm{D}$, at which it was
ejected from the component D and the epoch, $t_\mathrm{S1}$, when it
passed through the location of the stationary feature S1 (see Table~\ref{table:jet_kinematics}, 
Figs.~\ref{rfit} and \ref{flux-epoch}). For C12 component, these
epochs are estimated using the mean slope of the best-fits of
eight components (C4-C11).

\section{Link between subpc-scale jet and variable optical continuum emission}
\label{sec:link}
Optical fluxes and its errors measured at 5100\AA\ (rest frame) from observations of 3C\,390.3 between 1995 to 1999 (Fig.~\ref{flux-epoch}; red squares) are published in \cite{shapo} and \cite{sergeev}. During the time period from 2000 to 2007, spectra of 3C\,390.3 (97 nights) were taken with the 6\,m and 1\,m telescopes
of the Special Astrophysical Observatory of Russian Academy of Sciences (SAO RAS, Russia) and at the 2.1 m telescope of the ``Guillermo Haro Observatory'' at Cananea, Sonora (M\'exico), respectively. Long-slit spectrographs, equipped with CCD detector arrays, were used to cover the typical wavelength interval from
4000\,\AA\, to 7500\,\AA\, with spectral resolution between 5 \AA\ and 15 \AA\ and the signal-to-noise ratio $>$50 in the continuum near H$\alpha$ and H$\beta$ emission lines. The uncertainty in our flux determinations
for the continuum is $<$3~\%. Detailed description of the method and  measurements of optical fluxes are given in \cite{shapo}. Optical continuum fluxes measured at 5100 \AA\ between 2000 and 2007 are presented in the Table~\ref{table:optFlux} (for a full table see the online Supporting Information).

\begin{table}
\begin{center}
\caption{The first ten entities of optical continuum fluxes of 3C\,390.3 from the online electronic table. Columns: 1 -- UT date; 2 -- Julian date; 3 -- continuum flux at 5100~\AA\ (in units of 10$^{-15}$ erg s$^{-1}$cm$^{-2}$\AA$^{-1}$). \label{table:optFlux}}
\begin{tabular}{lcc}
\hline
  UT-date  &    JD       &  Flux at 5100 ~\AA   \\
           &  (2400000+) &                      \\
\hline
    1      &    2        &     3                \\
\hline
   2000Jul19 &   51745.363 &  1.557  $\pm$  0.047 \\
   2000Jul22 &   51748.230 &  1.602  $\pm$  0.048 \\
   2000Jul30 &   51756.293 &  1.543  $\pm$  0.046 \\
   2000Oct05 &   51823.141 &  1.410  $\pm$  0.042 \\
   2000Nov18 &   51867.137 &  1.558  $\pm$  0.047 \\
   2000Nov18 &   51867.223 &  1.661  $\pm$  0.050 \\
   2000Nov19 &   51868.129 &  1.660  $\pm$  0.050 \\
   2001Jan16 &   51925.633 &  1.533  $\pm$  0.046 \\
   2001Jan20 &   51929.621 &  1.465  $\pm$  0.044 \\
   2001Mar13 &   51981.949 &  1.396  $\pm$  0.042 \\
\end{tabular}
\end{center}

\end{table}

Variations of the optical continuum emission (Fig.~\ref{flux-epoch}; red squares) exhibit a superposition of weak flares (few weeks to several months), prominent flares (from months to years) and a long-term rising trend (few decades). The later is seen also in the historical light curve of the optical continuum in 3C\,390.3 \citep[e.g. ][]{shapo}.
It is remarkable, that all nine ejections are associated with nine optical flares happening on timescales from few months to few years (Fig.~\ref{flux-epoch}). This suggests that the frequency of optical flares and the ejection rate of jet components are interconnected. A link between the optical flares and the stationary component S1 in the radio jet is suggested by a correlation between the
maxima of optical flares and the characteristic epochs $t_\mathrm{S1}$ of the moving components C4--C12 (Fig.~\ref{flux-epoch}). For these nine components, the epochs $t_\mathrm{S1}$ of separation from the stationary feature S1 are coincident, within the errors, with the
maxima in the optical continuum flares. All nine
ejection events occur within $\sim 0.3$\,yr after a local maximum is
reached in the intensity of the optical continuum. Considering that there are about 14 years of optical data, we estimate the probability that maxima of optical flares and ejections from S1 are associated by chance in the 0.3 yr window. The probability of any single ejection event at S1 occurring randomly less than 0.3 yr after a maximum of optical flare is about 0.02.
The significance of the null hypothesis that all nine events happen by chance is $\approx10^{-10}$ and can be rejected at a confidence level of $>99.99$\,\% (for a description of the statistical test see Marscher et al. 2002). The significance is still high ($\approx 10^{-9}$) even without two components C4 and C8 for which the gaps in the optical data do not allow the timing of the optical maxima although the data is consistent with a maximum.

This is a clear evidence that the passage of moving radio knots through the location of the stationary component S1 and optical continuum flares are physically related. The radio events follow the maxima of optical flares with the mean time delay of $(\approx 0.10\pm 0.04)$\,yr. The average distance between the region where the optical flare reaches maximum and the component S1 is measured to be $0.04$ pc (or 0.36 pc from D) adopting that the knots propagate with the mean apparent speed $\approx c$ during the time period of $\simeq 0.1$ yr. Radio events are associated with optical flares occurring on time-scales of about a few months to a few years (Fig.~\ref{flux-epoch}). The rise of all nine optical flares occurs between two radio events, the ejection from component D and separation from S1 (Fig.~\ref{flux-epoch}) suggesting that the generation of variable optical emission happens in the innermost part of the jet, between D and S1, as a result of propagation of these ejections along the jet. The generation of optical flares and its characteristics such as the amplitude, timescale and the frequency are likely to be related to the properties of the subparsec-scale jet: ejection rate, structure and kinematics. Therefore, understanding the physical nature of the regions D and S1 is important for understanding the structure of the central engine in 3C\,390.3 and location of continuum source(s) producing the variable continuum radiation.

\section{Variable radio emission of the stationary components in the jet}
\label{sec:var_emission}
Variations of the flux densities of D and S1 components of the jet
($f_{\rm D}$ and $f_{\rm S1}$) are shown in Fig.~\ref{flux-epoch}
(cyan circles and cyan diamonds). The relatively sparse time sampling of the VLBI measurements does not allow us to test for
a correlation between radio flares and epochs of ejections. Fractional
variability amplitude of S1 ($0.44\pm0.18$), defined as in \cite{lu01}, is about two times higher than that of D and the optical emission ($0.3\pm0.17$ and $0.24\pm0.07$, respectively). The optical
and radio flux densities of D increase by the same factor ($\approx$3 times)
from 1996 to 2008 showing a surprisingly similar behaviour over a time period of 14 years, while the radio flares from S1, varying on scales from a
few months to a few years, show some correlation with the optical flares having
comparable timescales. Moreover, magnitudes of the optical and radio flares
are comparable in S1, with the flux density varying by a factor of three
for optical continuum flux and a factor of four in the radio.
It is therefore likely that the optical variations on timescales of few decades are
related to the radio variability of the component D, while
the optical flares are coupled with radio flares of S1 component of
the jet.

\begin{figure}
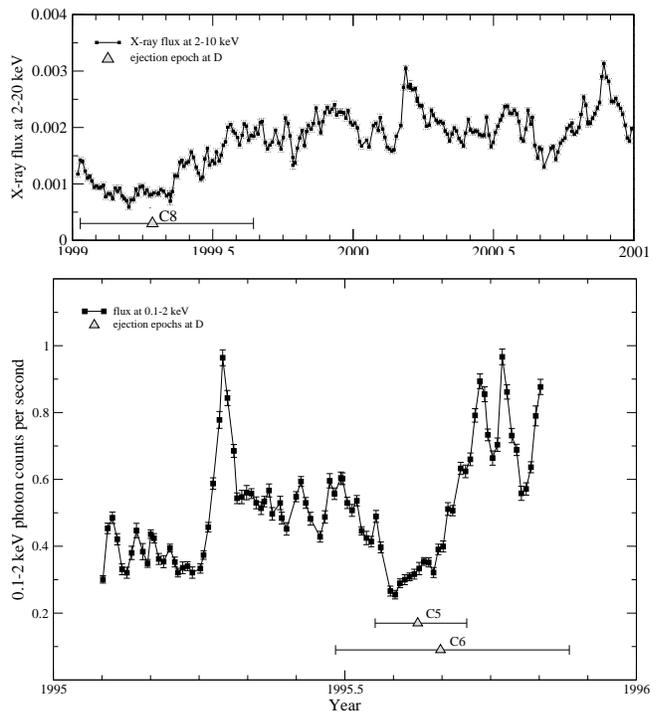

\includegraphics[width=0.48\textwidth]{f4a.eps}
\includegraphics[width=0.48\textwidth]{f4b.eps}
  \caption {X-ray flux variations in 3C\,390.3 and the epoch of jet
  component ejection at D, the base of the jet. \emph{Top}: Hard X-ray
  fluxes at 2 keV to 20 keV (Gliozzi et al. 2006) and the epoch of
  ejection of C8. \emph{Bottom}: Soft X-ray fluxes at 0.1 keV to 2 keV
  (Leighly et al. 1997) and the epochs of ejection of C5 and
  C6. 1$\sigma$ error bars are presented for all data.} \label{xray}
\end{figure}

We used the Z-transformed correlation function \citep{alexander97} to
calculate the correlation between $f_{\rm opt}$ and $f_{\rm S1}$ as this
method can deal with sparsely sampled data. The correlation function
between $f_{\rm opt}$ and $f_{\rm S1}$ have two peaks at a timelag  around
$-50^{+8}_{-7}$ days and $-350^{+22}_{-15}$ days (implying that radio flares
lead optical flares with time lag of about one year), with a correlation
coefficient of around $0.55^{+0.26}_{-0.22}$. Correlation function between $f_{\rm opt}$ and $f_{\rm D}$ has a flat shape and high correlation coefficient, around $0.85^{+0.10}_{-0.07}$, with time lags from 200 days to 400 days and from -400 days to -500 days. The present data suggest a correlation between variations of optical and radio emission curves but uncertainties of the correlation coefficient are large (because of the poor radio sampling) for making final conclusions. A denser VLBI radio sampling covering a time period of several years is needed to improve the correlation observed between the radio and optical flares.

\section{Physical identification of the regions D and S1}
\label{sec:id}
The presence of several apparently stationary features in the radio jet
of 3C\,390.3 may be explained in several different ways.
According to present models of relativistic jets
\citep[][ and references therein]{marscher08}, the plasma material accreted onto central nucleus is collimated into the jet. At the base of the jet, the jet flow either becomes supersonic \citep{daly88} or optically thin
\citep{konigl81,lobanov98} or releases the energy contained in the
Poynting flux \citep{romanova96}. The plasma material is then accelerated
and collimated into continuous jet flow by helical magnetic fields on typical scales of $\sim 10^4$ Schwarzschild radii. At the end of the acceleration and collimation zone ($\la 1$ pc), the jet flow may become turbulent (due to weakening of the helical
magnetic field) and end in a standing conical shock associated with the radio core of the jet \citep{marscher08}.

The link between the ejection epochs of jet components and the dip in the X-ray was found for two radio galaxies 3C\,120 and 3C\,111 \citep{marscher02,marscher06}. This correlation was interpreted as accretion of the X-ray-emitting gas in the disk into the central black holes and ejection of a fraction of the infalling matter into the jet \citep{marscher02}. We found a similar correlation between the ejection epochs of the components C5 and C8 and the dip in the X-ray flux (see Fig.~\ref{xray}) and hardening of the spectrum \citep{arshakian06}. The ejection of C4 and C8 components happens during the dip in the X-ray flux suggesting that the component D is located near the black hole. Adopting an angle\footnote{On the assumption that the pattern speed and bulk speed of the jet of 3C\,390.3 are equal, the jet inclination angle
$\theta\sim50^{\circ}$, bulk Lorentz factor $\gamma\sim2$ and
beaming angle $\psi \approx \gamma^{-1}\sim30^{\circ}$ are estimated
using the variable Doppler factor $\delta=1.16$ \citep{lahteenmaki}
and the maximum apparent speed of $1.6\,c$ observed in the compact
jet.} of 50$^\circ$ between the jet and the line of the sight, the
corresponding de-projected distance between D and S1 is
$0.3$\,pc$/\sin50^{\circ} \approx 0.4$\,pc.
The average brightness temperature of the component D, $T_{\rm b}(D)\sim 1.2\times10^{11}$\,K, is close to the inverse Compton limit $\sim 5 \times 10^{10}$\,K \citep{ken69} while the $T_{\rm b}(S1)\sim 2.2\times10^{10}$\,K of S1 component suggests that the synchrotron radiation mechanism maintains the approximate equipartition in energy between the particles and the magnetic field in the region around S1 \citep{readhead94}.
The component D can be identified with the accretion disk or hot corona, actual base of the jet or the radio core.

The high brightness temperature rules out the D being the accretion disk or hot corona since the brightness temperature for three major accretion disk models is estimated to be less than $10^{10}$\,K. \emph{Standard disks} are characterized by a balance between radiative cooling and viscous heating \citep{shakura73}. For a typical quasar with $M_{\rm bh} = 10^8 M_{\sun}$ and $L \sim 10^{38}$ W, the effective
temperature will be $T_{\rm eff} \sim 10^5$\,K at a radius of $r\sim 100 r_{\rm s}$ ($r_\mathrm{s} = 2\,G\,M_\mathrm{bh}/c^2$
is the Schwarzschild radius for a black hole of mass
$M_\mathrm{bh}$, where $G$ is the Newtonian gravitational constant). However, a hot thin corona ($kT_{\rm e} \sim 70-250$ keV; $T_{\rm e} \sim 0.8 - 3\times 10^9$\,K) above the disk layer can be a bright source of radiation by means of inverse Compton scattering of UV photons from the disk \citep{haardt91,haardt93}.
\emph{Advection-dominated accretion flow} (ADAF) is radiatively inefficient so that most of heat produced through
accretion is preserved \citep{narayan94}. The ADAF model predicts that the mass accretion rate is a factor of $\sim 10^{-4}$ of the Eddington limit and that the electron temperature is above $10^9$\,K for $r < 100r_{\rm s}$ \citep{manmoto97}. The \emph{slim disk model}, with an extremely high mass accretion rate above the Eddington limit, was
proposed by \cite{abramowicz88}. The disk is hotter than the standard disk because of the high accretion rate. The brightness of the blackbody radiation from slim disks is $\sim 10^6$\,K. However, as in the case of standard disks, a hot corona may produce a radiation with $T_{\rm e} \sim 0.8 - 3\times 10^9$\,K.

The mean brightness temperature of the D component is higher than that of the S1, $T_{\rm b}(D)>\, T_{\rm b} (S1)$, suggesting that D is the radio core of the jet rather than the base of the jet. Were D the core of the counterjet and S1 the core of the approaching jet, we should expect that $f_{\rm D}<f_{\rm S1}$ because of relativistic (Doppler) beaming effect. The fact that $f_{\rm D}>f_{\rm S1}$ over the 14 years of monitoring period (see Fig.~\ref{flux-epoch}) rules out D being the core of the counterjet. The component D should then be associated with the radio core of the jet and its emission is non-thermal and generated by inverse Compton mechanism ($T_{\rm b}\sim 1.2\times10^{11}$\,K). Then the component S1 should be associated with the standing shock (a common feature seen in superluminal AGN) downstream from the radio core. This is supported by the average $T_{\rm b} \sim 2.2\times 10^{10}$\,K estimated for S1 component, which is close to the equipartition limit ($\sim 5\times 10^{10}$\,K; Readhead 1994).

Alternative scenario assumes that the component D is the base of the jet located in the immediate vicinity of the central black hole. Then the S1 stationary feature should be the radio core of the jet.  Location of D near the accretion disk is supported from the X-ray data (Fig.~\ref{xray}). Very long-term variations (on time-scales of decades) of radio emission from D component is not characteristic for the jet emission which varies on time scales from months to years. Variations from D component on time-scales of decades mimic very long-term variations of optical continuum emission (Fig.~\ref{flux-epoch}) which is probably generated in or above a hot corona at a distance $\sim$200\,$R_\mathrm{s}$ above the accretion disk \citep{fabian04,ponti04}. The non-thermal radio emission at the base of the jet can be generated by synchrotron or inverse Compton ($T_{\rm b}(D)\sim 1.2\times10^{11}$\,K) mechanisms by free electrons of corona accelerated in the rotating and twisted magnetic fields generated by the accretion disk.

Another possibility is that D and S1 are two independent, active nuclei in the binary black hole system. The binary black hole scenario cannot explain
the correlated, simultaneous variability of the red and blue wings
of the H$\beta$ line \citep{shapo}. If the two black holes are
located in D and S1, the total mass of the binary must exceed $\sim
5\times 10^9\,(P_\mathrm{obs}/1000\,\mathrm{yr})\,M_\odot$.
The observed radial velocity changes indicate a possible periodicity shorter
than 100\,yr \citep{shapo}, pushing the total mass of the binary to $\sim
10^{11}\,M_\odot$, in contradiction with the mass of $\sim4\times
10^{8}\,M_\odot$ measured using the reverberation mapping
technique \citep{kaspi00}. The trends observed in the radial
velocities cannot be reconciled with the orbital motion in a binary
black hole \citep{eracleous97}. Thus it is also difficult to use a binary 
system with two black holes located in D and S1 as an explanation for the
observed properties of 3C\,390.3.

To understand the radiation mechanism and structure of the inner
nuclear region of radio-loud galaxy 3C\,390.3, it is fundamental to identify and locate 
the D and S1 regions. This requires 
multi-frequency VLBI observations to measure the spectral index and
polarization of stationary components of the compact jet.

\section[]{Discussion and conclusions}
\label{sec:discussion}

Soldi et al. (2008) have carried out multiwavelength of the quasar 3C\,273 \citep{soldi08}, and have shown that its emission is complex due to the presence of many emission regions in the central engine. They suggested that variable optical emission is non-thermal and it is likely to be generated in the base of the jet. Evidence for optical synchrotron emission generated in and near the radio core of the jet comes from monitoring of radio and optical polarized emission of the quasar PKS\,0420-014 \citep{darcangelo07}. Concurrent changes of the polarization angle at optical and radio (43\,GHz) indicate that these variable emission is non-thermal and originate in the same region around the radio core. While the emission of the accretion disk is viewed directly for a quasar, it is hidden in radio galaxies by the dusty torus. Very weak (or absent) blue bump in 3C\,390.3 \citep{wamsteker} indicates that the accretion disk is obscured and most of variable optical continuum emission could be generated in the jet.

The significant correlation found between the optical flares and the epochs $t_{\rm S1}$ can be interpreted in terms of disturbances (radio knots) propagating from the location of D component to the S1 stationary component. If the component D is the base of the jet then the following physical scenario is likely. A portion of the disk X-ray emitting material accreted onto central nucleus is ejected from the base of the jet (in the immediate vicinity from the black hole) at the epoch $t_{\rm D}$. The knot of electrons accelerated to high energies
radiates a rising emission in X-ray, optical and radio as it propagates upstream from the base of the jet. The jet becomes visible in radio at $t_{\rm S1}$ because of the emitting region of the jet becoming optically thin or denser \citep{konigl81,romanova96}. Alternatively, if the component D is the core of the jet then the rising synchrotron emission from X-ray to radio bands can be produced between D and S1 components by acceleration of the knot (shock?) in a helical magnetic field \citep{marscher08} and/or evolution of the shock spectrum (or turnover frequency; Valtaoja et al. 1992). The minimum Doppler factor $D_{\rm min}=1.5$ is required to increase the optical flux by factor of 2.5 (see Fig.~\ref{flux-epoch}), if the jet flow accelerates from $\gamma=1$ to $\gamma=2$ and the apparent speed of the jet is $1.6c$. This value $D_{\rm min}=1.5$ is in reasonably good agreement with the measured variable Doppler factor $\approx1.2$ \citep{lahteenmaki}. Then the component S1 should be a standing shock formed by continuous relativistic flow \citep{gomez95}. The moving knot would be compressed by the shock wave thus producing variable emission in a wide range of electromagnetic spectrum as indicated from multiwaveband polarization variability \citep{darcangelo07}. As the disturbance passes the standing shock S1, which manifests the end of the acceleration zone, it expands adiabatically because of low magnetic field strength and low electron density, and the optical flux drops down on time scales of $\la 1$\,yr, while the high frequency radio emission lasts much longer because low-energy electrons have longer lifetimes. In this scenario, the timescale and amplitude of non-thermal optical flares depend on kinematics and emitting power of moving perturbations in the innermost subparsec-scale region of the jet.

Variations of the broad-line emission in 3C\,390.3 follow the optical continuum with some time delay which is estimated by various authors to be in the range from about 20 days to 100 days \citep{wandel99,kaspi00,shapo,sergeev}. If the variable optical emission is being generated by shocks in the innermost part of the jet then the beamed continuum emission from the jet must ionize the gas around the jet thus producing the BLR downstream from the `classical' virialized BLR ionized by the optical emission generated in the accretion disk. An outflowing, non-virialized BLR can be generated in the rotating subrelativistic outflow \citep{murray97,proga00} surrounding the jet. 
In BL Lac type objects (in which the jet oriented near the line-of-sight) moving superluminal
components also may produce the short-term radio flares, optical, X-ray and
even gamma-ray flares \citep{marscher08} when passing near the line-of-sight of the observer. In the case of 3C\,390.3 the jet is inclined at $\approx
50^{\circ}$, hence, the Doppler beaming of continuum emission is weak towards the observer but the emission is enhanced in the direction of the jet and able to ionize the conical outflowing BLR with a half-opening angle$^2$ $\psi \approx \gamma^{-1}\sim30^{\circ}$.
In radio-loud galaxies there could be two regions of generation of variable optical emission, non-thermal emission in the innermost part of the jet and thermal optical emission in the accretion disk.

The existence of the jet-excited outflowing BLR in 3C\,390.3 will question the assumption of virilized motion in the BLR \citep{kaspi00} of all
radio-loud AGN, galaxies and quasars, and, hence, the applicability
of the reverberation mapping \citep{peterson02} to estimate the
black hole masses of radio-loud AGN. Time delays and profile widths
measured during periods when the jet emission is dominant may not
necessarily reflect the Keplerian motion in the disk, but rather
trace the rotation and outward motion in an outflow. This can result
in large errors in estimates of black hole masses made from
monitoring of the broad-line emission.  In the case of 3C\,390.3,
the black hole mass ($2.1\times10^9\,M_{\odot}$)
estimated effectively from the measurements near the maximum in the
continuum light curve \citep{shapo} is significantly larger than the
values ($(3.5\,\mathrm{to}\,4)\times 10^8\,M_{\odot}$) reported in other works
\citep{wandel99,kaspi00}. This difference is reconciled by
considering the line width ($v_{\rm FWHM}\approx 10500$\,km\,s$^{-1}$) and the time delay ($\tau \approx 24$ days) between the optical
continuum and line fluxes near the minimum of the continuum light
curve, which yields $M_\mathrm{bh} \approx 1.45 \times 10^{5}\,M_{\sun} (c \tau/{\rm lt-day}) v^{2}_{\rm FWHM}= 3.8\times 10^8\,M_{\odot}$. The
possible existence of an outflow-like region in a number of
radio-loud AGN should be taken into account when estimates of the
nuclear mass are made from the variability of broad emission lines.

The presence of the jet-excited nonvirial BLR in radio-loud AGN
is capable of explaining some of the spectral characteristics of
emission lines. Depending on
the orientation of the jet, the approaching and rotating outflow
material in the BLR will imprint prominent signatures on the emission
lines. At small viewing angles of the jet this BLR may produce
blue-shifted and single-peaked broad emission lines, while non-shifted
and double-peak emission lines \citep{eracleous03} will be observed at
large angles of the outflowing BLR to the line of sight. The narrow emission lines will have similar characteristics \citep{boroson05} being ionized in the approaching subrelativistic outflow by the beamed
continuum emission of the jet.

The correlation found for 3C\,390.3, between the epochs of maxima in optical flares and passing of radio knots through S1 component, is also confirmed for the radio galaxy 3C\,120 \citep{tavares09}. To check whether this correlation is common for other galaxies we conducted coordinated long-term radio-optical observations of several nearby radio-loud galaxies. A denser VLBI radio sampling covering a time period of several years is required to resolve the correlation between variabilities of the radio emission of the jet components and the optical continuum, and further to constrain the models for the nuclear region of radio-loud galaxies. \\

The principal results of this work comes from analysis of combined
radio VLBI (15\,GHz), optical/UV and X-ray data of the
radio-loud galaxy 3C\,390.3, and they can be summarized as follows:
\begin{enumerate}
 \item Structure of the parsec-scale jet: from ten epochs of VLBI
 observations we identified three stationary components (D, S1 and S2)
 and nine moving components (C4-C12) of the jet on scales of a few
 parsecs. Apparent speeds of moving components are estimated to be in
 the range from 0.7\,$c$ to 1.6\,$c$.

 \item  We found a new correlation between the local maxima in the optical
 continuum light curve and the epochs at which the moving components
 of the jet pass the stationary radio feature S1. Nine radio events follow the peaks of optical flares with the mean time delay of $1.2\pm 0.4$ months. Optical flares brighten when radio knots move between stationary components D and S1.

 \item Identification and location of D and S1 regions: analysis of
 available archival X-ray monitoring data and ejection epochs of two components, C5 and C8, revealed that the ejection of these components from D occurs during the dip in the X-ray emission. It is most likely that the component D is associated with the VLBI core of the jet (rather than the base of the jet) and the component S1 is a standing shock located at a distance of $\approx0.4$ pc downstream from the component D.

 \item These results have important implications for the structure of the
 sub-parsec-scale nuclear region of the radio-loud galaxy 3C\,390.3. We suggest that the relativistic plasma of the jet is a dominant source of variable optical continuum emission: the bulk of optical continuum optical emission on timescales from few months to years are likely to be generated by knots of high-energy electrons propagating in the innermost 0.4\,pc region of the jet between stationary components D and S1. In this scenario, the timescale, amplitude and frequency of optical synchrotron flares depend on energetics, kinematics and rate of ejected radio knots. The beamed continuum emission from the jet ionizes a gas in a subrelativistic outflow along the jet, which results in a formation of two non-virialized outflowing BLRs along the jet and counterjet.

\end{enumerate}

\section*{Acknowledgments}
We acknowledge helpful discussions with N.G. Bochkarev, thank
E. Valtaoja for valuable comments and discussions which significantly improved the paper, and A. Roy for useful discussions and careful reading of the draft of
the paper. This work was supported by grants form INTAS
(grant N96-0328), RFBR (grants N97-02-17625 N00-02-16272,
N03-02-17123 and 06-02-16843), State program 'Astronomy' (Russia), and
CONACYT research grant 54480 (Mexico).
JLT acknowledges support from the CONACyT program for PhD studies,
the  International Max-Planck Research School for Radio and
Infrared Astronomy at the Universities of Bonn and Cologne  and the
Deutscher Akademischer Austausch Dienst for a short-term
scholarship in Germany. This research has made use of data from the
MOJAVE database that is maintained by the MOJAVE team \citep{lister09}. 
The National Radio Astronomy
Observatory is a facility of the National Science Foundation operated
under cooperative agreement by Associated Universities, Inc.

\bibliographystyle{mn2e}

\appendix

\newpage

\section[]{Properties of the VLBA images and model fits of 3C 390.3}
\label{app:A}

\begin{table*}
\begin{center}
\caption{Properties of the VLBA images and model fits of 3C\,390.3.
\label{tbl-1}}
\begin{tabular}{l rr rr rr rr rr}
\hline\hline
 & \multicolumn{2}{|c|}{$S_\mathrm{total}$} &
 \multicolumn{2}{|c|}{$S_\mathrm{peak}$} &
 \multicolumn{2}{|c|}{$S_\mathrm{min}$} &
 \multicolumn{2}{|c|}{$\chi^2$} &
 \multicolumn{2}{|c}{$\sigma_\mathrm{rms,uv}$} \\ \cline{2-11}
\raisebox{2.4ex}[0cm][0cm]{Epoch} &
 \multicolumn{1}{c}{I} & \multicolumn{1}{c|}{M} &
 \multicolumn{1}{c}{I} & \multicolumn{1}{c|}{M} &
 \multicolumn{1}{c}{I} & \multicolumn{1}{c|}{M} &
 \multicolumn{1}{c}{I} & \multicolumn{1}{c|}{M} &
 \multicolumn{1}{c}{I} & \multicolumn{1}{c}{M} \\
\hline
1994.67 & 423.9 &422.9 &213.3 &211.1 &$-2.2$  &$-3.4$ &1.175 &1.186 &253.1 &253.8 \\
1995.96 & 463.1 &466.2 &190.8 &190.8 &$-1.6$  &$-2.8$ &0.328 &0.334 &119.2 &120.3 \\
1996.37 & 471.6 &477.6 &192.4 &192.6 &$-1.3$  &$-3.0$ &0.453 &0.475 & 83.4 & 85.5 \\
1996.82 & 389.5 &399.1 &174.2 &174.2 &$-1.5$  &$-3.2$ &1.235 &1.504 & 42.5 & 45.2 \\
1997.19 & 370.0 &377.5 &179.1 &180.4 &$-0.7$  &$-2.2$ &1.533 &1.800 & 39.9 & 43.0 \\
1997.66 & 418.3 &425.2 &267.4 &267.2 &$-1.8$  &$-2.2$ &0.575 &0.580 &152.1 &152.5 \\
1998.21 & 392.4 &389.6 &191.2 &191.4 &$-1.1$  &$-1.3$ &0.429 &0.430 & 94.6 & 94.7 \\
1999.85 & 313.2 &315.8 &216.9 &218.6 &$-1.1$  &$-1.3$ &1.103 &1.149 & 31.0 & 31.4 \\
2000.65 & 310.4 &300.2 &161.8 &167.4 &$-5.3$  &$-7.3$ &1.234 &1.595 & 38.9 & 42.2 \\
2001.17 & 270.6 &280.1 &209.4 &207.6 &$-1.2$  &$-2.6$ &1.250 &1.578 & 32.2 & 35.4 \\
2002.13 & 301.3 &297.6 &217.0 &217.3 &$-0.9$  &$-1.8$ &1.094 &1.437 & 28.4 & 31.6 \\
2003.02 & 268.1 &254.3 &211.0 &212.5 &$-3.3$  &$-5.3$ &1.275 &1.929 & 35.0 & 41.2 \\
2005.13 & 299.6 &303.9 &220.3 &217.0 &$-3.7$  &$-2.8$ &1.013 &0.996 & 26.1 & 25.5 \\
2006.11 & 344.8 &337.9 &295.9 &287.3 &$-1.7$  &$-3.1$ &1.074 &1.446 & 91.6 & 54.7 \\
2006.31 & 340.6 &348.3 &278.8 &279.0 &$-3.1$  &$-3.4$ &1.127 &1.228 & 31.6 & 32.5 \\
2006.54 & 390.4 &391.2 &306.5 &305.3 &$-3.0$  &$-3.5$ &1.049 &1.092 & 33.3 & 33.7 \\
2006.75 & 379.2 &377.6 &303.0 &303.8 &$-3.2$  &$-2.8$ &1.052 &1.067 & 31.6 & 31.9 \\
2007.01 & 405.1 &409.8 &350.5 &350.3 &$-1.6$  &$-1.8$ &1.109 &1.109 &118.9 &118.8 \\
2007.63 & 444.5 &445.7 &380.6 &380.6 &$-1.6$  &$-1.8$ &1.131 &1.141 & 76.3 & 76.3 \\
2008.41 & 468.6 &462.6 &394.3 &394.4 &$-2.1$  &$-2.7$ &1.189 &1.213 & 68.7 & 68.5 \\
2008.60 & 429.4 &422.6 &344.0 &341.5 &$-1.9$  &$-3.1$ &1.138 &1.267 & 65.8 & 69.2 \\

$<\mathrm{I/M}>$ & \multicolumn{2}{|c|}{0.999} &
\multicolumn{2}{|c|}{1.000} &
\multicolumn{2}{|c|}{0.7242} &
\multicolumn{2}{|c|}{0.906} &
\multicolumn{2}{|c}{0.999}  \\
\hline
\end{tabular}
\end{center}
\medskip
\begin{flushleft}
Column designation: $S_\mathrm{total}$ -- total flux in the image
(``I'') and in the respective Gaussian (``M'') component models,
$S_\mathrm{peak}$ -- peak flux density in the respective images and
models, $S_\mathrm{min}$ -- minimum flux density in the respective
images and models, $\chi^2$ -- goodness-of-fit parameter of the
respective images and models, $\sigma_\mathrm{rms}$ -- r.m.s
visibility noise for the respective images and models.
\end{flushleft}
\end{table*}

Table~\ref{tbl-1} compares the CLEAN component models (denoted ``I'')
and Gaussian model fits (denoted ``M'') for the VLBA data used in the
paper. The columns are: $S_\mathrm{total}$ -- total flux density
[mJy/beam]; $S_\mathrm{peak}$ -- peak flux density [mJy/beam];
$S_\mathrm{min}$ -- minimum flux density [mJy/beam]; $\chi^2$ --
goodness of the fit parameter; $\sigma_\mathrm{rms,uv}$ --
root-mean-square between the observed and model visibilities
[mJy]. Last row presents average ratios between the respective image
and model fit properties. The ratios are close to unity for the total
and peak flux densities. The $\chi^2$ parameter and the visibility
r.m.s. are only slightly larger for the Gaussian model fits, which
indicates that the model fits represent the structure adequately. The
higher maximum negative flux density in the model fits (column
$S_\mathrm{min}$) indicates that that the SNR of the Gaussian fits is
on average 1.5 times lower than that of the VLBI images (due to
increased non-Gaussian shapes of low-brightness regions). This
reduction does not affect the fitted values of the component
parameters (albeit it does increase the parameter errors) as it is
related to extended emission associated with the underlying flow.

Properties of individual emitting regions in the radio jet are
determined by fitting Gaussian patterns to interferometric amplitudes
and phase closures. Errors of the modelfits are estimated by varying
the best fit parameters and determining 1-$\sigma$ confidence limits
from the resulting $\chi^2$ distribution. This method accounts for
correlations between parameters of adjacent modelfit components and
provides conservative error estimates.  Fidelity of the component
sizes and separations obtained from the modelfits is further examined
by comparing them with resolution limits calculated from the
SNR of the detections. The resolution limit, $\theta_\mathrm{min}$,
can be estimated for a Gaussian modelfit component from
\[
\theta_\mathrm{min} =
\left[\frac{16\, \ln\,2}{\pi}  \ln\left(\frac{S}{S-1}\right) b_\mathrm{maj}\,b_\mathrm{min}\right]^{1/2}\,,
\]
where $S$ is the signal-to-noise of component detection and
$b_\mathrm{maj}$, $b_\mathrm{min}$ are the major and minor axis of the
point-spread function derived from the interferometric coverage of the
data.  The limiting separation, $r_\mathrm{min}$, at which two
Gaussian components with sizes $\theta_{1,2}$ can be fitted adequately
to visibility data can be estimated from
\[
r_\mathrm{min} = \frac{1}{2}
\frac{\theta_\mathrm{min,1}\, \theta_\mathrm{min,2}}
{(b_\mathrm{maj}\,b_\mathrm{min})^{1/2}} \left[
\left(\frac{\theta_1}{\theta_\mathrm{min,1}}\right)^2 +
\left(\frac{\theta_2}{\theta_\mathrm{min,2}}\right)^2
\right]^{1/2}\,,
\]
where $\theta_\mathrm{min,1}$ and $\theta_\mathrm{min,2}$ are the
minimum resolvable sizes for the two components.  Table~\ref{tbl-2}
lists the ratios between the measured sizes of and separations of the
components D and S1 and their respective limiting values. For resolved
components the ratios between the measured sizes $\theta$ and their
respective $\theta_\mathrm{min}$ are larger than unity.  For the
component separations, ratios large than unity imply that the
components parameters can be reliably estimated from the modelfit.

The SNR of detection varies from 30 to 900 for S1 and from 100 to 860
for D and the corresponding resolution limits are smaller than 0.2 mas
for both features. The ratios between measured and limiting sizes and
separations of D and S1 indicate that only the size measurements made
for the first data point for S1 and the last one for D are coming close
to the resolution limits, while all other measurements are certainly
above these limits, both for the component sizes and separations.  The
limiting ratios and separations can be used to estimate the
probability of the component flux to be correlated:
\[
p_\mathrm{c}(r) = \left(\frac{r_\mathrm{min}}{r}\right)^2
\left[
\left(\frac{\theta_1}{\theta_\mathrm{min,1}}\right)^2 +
\left(\frac{\theta_2}{\theta_\mathrm{min,2}}\right)^2
\right] \ln\left(\frac{S_1}{S_1-1}\right)\times \]
\[\ln\left(\frac{S_2}{S_2-1}\right) \left[\ln\left(\frac{S_1}{S_1-1}\right)^2 + \ln\left(\frac{S_2}{S_2-1}\right)^2  \right]^{-1}\,.
\]
This formula gives $p_\mathrm{c} \equiv 1$ for two identical Gaussian
features separated by $r_\mathrm{min}$.  The correlation probabilities
(listed in the last column of Table~\ref{tbl-2}) are smaller than 6\,\%
for all of the model fits, implying that the flux densities of the
component D and S1 estimated from these modelfits are uncorrelated.

\begin{table}
\begin{center}
\caption{Resolution limits and fidelity of the flux densities in the modelfits
\label{tbl-2}}
\centering
\begin{tabular}{l|cc|cc|r|r}
\hline\hline
      & \multicolumn{2}{c|}{SNR} & \multicolumn{2}{c|}{$\theta/\theta_\mathrm{min}$} & $r/r_\mathrm{min}$ & \multicolumn{1}{c}{~} \\\cline{2-6}
\raisebox{2.4ex}[0cm][0cm]{Epoch}
       & D & S1 &  D  &    S1   &    \multicolumn{1}{c|}{D--S1} &
\multicolumn{1}{c}{\raisebox{2.4ex}[0cm][0cm]{$p_\mathrm{c}$}}\\ \hline
1994.67 & 433.3 &  32.0 &  3.13 & 0.99 &   3.8  & 0.0540 \\
1995.96 & 454.3 & 303.9 &  1.90 & 1.64 &  17.6  & 0.0094 \\
1996.37 & 388.8 & 116.4 &  1.51 & 1.79 &   9.1  & 0.0182 \\
1996.82 & 481.6 & 66.3  &  2.71 & 2.32 &   5.3  & 0.0608 \\
1997.19 & 861.8 & 130.8 &  3.16 & 1.67 &  11.0  & 0.0156 \\
1997.66 & 475.2 & 140.2 &  1.53 & 1.76 &  11.3  & 0.0115 \\
1998.21 & 535.1 & 407.3 &  2.62 & 2.88 &  13.5  & 0.0401 \\
1999.85 & 568.4 & 266.0 &  2.81 & 1.31 &  13.4  & 0.0205 \\

2000.65 & 115.9 & 70.4 &1.08 &1.58 & 6.1 &0.0431 \\
2001.17 & 508.5 &239.4 &1.06 &1.41 & 9.7 &0.0127 \\
2003.02 & 632.4 &356.2 &1.21 &1.15 & 6.6 &0.0272 \\
2005.13 & 406.2 & 38.4 &2.42 &0.99 & 3.8 &0.0428 \\
2006.11 & 499.7 & 90.8 &1.24 &1.19 &10.6 &0.0046 \\
2006.31 & 455.4 &116.2 &1.29 &1.65 & 9.3 &0.0122 \\
2006.54 & 681.9 &265.7 &1.71 &1.91 & 9.4 &0.0249 \\
2006.75 & 573.2 &174.6 &1.21 &1.54 &11.0 &0.0088 \\
2007.01 & 688.3 &128.4 &1.62 &1.34 & 9.0 &0.0097 \\
2007.63 &1069.1 &351.5 &1.28 &1.24 &13.3 &0.0053 \\
2008.41 &1599.0 &423.5 &1.89 &1.02 &10.7 &0.0099 \\
2008.60 &1633.3 &266.2 &2.39 &1.39 & 7.0 &0.0245 \\\hline

\end{tabular}
\end{center}
{Column designation: SNR -- signal-to-noise of component
detection, $\theta/\theta_\mathrm{min}$ -- component ratio of the
measured and minimum resolvable sizes, $r/r_\mathrm{min}$ -- ratio of
the measured distances the minimum resolvable separations between the
components D and S1, $p_\mathrm{c}$ -- probability of the measured flux
densities of D and S1 being correlated.}
\end{table}

\bsp

\label{lastpage}

\end{document}